%% file: main.tex
\documentclass{article}
\usepackage{spconf,amsmath,amsfonts,graphicx}
\usepackage{amsmath,amsfonts,tabularx,multirow}
\usepackage{amssymb,textcomp,mathtools}
\usepackage{color}
\usepackage{algorithm} 
\usepackage{algorithmic}  
\usepackage[algo2e]{algorithm2e} 
\usepackage{bm,mathtools,upgreek}
\usepackage{verbatim, cite}
\usepackage{bm,upgreek,algorithm,hyperref}
\usepackage{multirow, booktabs, hhline, array}
\usepackage{cite, url, makecell, setspace}
\usepackage{acronym}
\usepackage{xcolor}

\def\thline{\noalign{\hrule height 1.0pt}}

\renewcommand{\vec}[1]{\bm{\mathrm{#1}}}
\title{Continuous speech separation using speaker inventory for long multi-talker recording}
\name{\begin{tabular}{c}Cong~Han$^1$, Yi~Luo$^1$, Chenda~Li$^2$, Tianyan~Zhou$^3$, Keisuke~Kinoshita$^4$,\\
Shinji~Watanabe$^5$, Marc~Delcroix$^4$, Hakan~Erdogan$^6$, John R. Hershey$^6$, Nima~Mesgarani$^1$, Zhuo~Chen$^3$\end{tabular}}

\address{
    $^1$Columbia University,
    $^2$Shanghai Jiao Tong University,
    $^3$Microsoft Corporation, \\
    $^4$NTT Corporation,
    $^5$Johns Hopkins University,
    $^6$Google Research
}

\begin{document}
\ninept
\maketitle
\setlength{\abovedisplayskip}{5pt}
\setlength{\belowdisplayskip}{5pt}
\setlength{\abovedisplayshortskip}{5pt}
\setlength{\belowdisplayshortskip}{5pt}
\begin{abstract}
\input{abstract}
\end{abstract}
\begin{keywords}
Speech separation, continuous speech separation, speaker inventory, embedding clustering
\end{keywords}
\section{Introduction}
\label{sec:intro}
\input{introduction}

\vspace{-0.11cm}

\section{SSUSI using pre-enrolled utterance}
\label{sec:s2}
\input{ssusi}

\vspace{-0.11cm}

\section{Continuous SSUSI using self-informed mechanism for inventory construction}
\label{sec:s3}
\input{cssusi}

\vspace{-0.11cm}

\section{Experimental settings}
\label{sec:exp}
\input{experiments}

\vspace{-0.11cm}

\section{Results and discussions}
\label{sec:results}
\input{results}

\vspace{-0.11cm}

\section{Conclusion}
\label{sec:con}
\input{conclusion}
\vspace{-0.11cm}

\section{Acknowledgement}
\label{sec:ack}
The work reported here was started at JSALT 2020 at JHU, with support from Microsoft, Amazon and Google.

\footnotesize 
\bibliographystyle{IEEEbib}
\bibliography{main}

\end{document}

%% file: abstract.tex
Leveraging additional speaker information to facilitate speech separation has received increasing attention in recent years. Recent research includes extracting target speech by using the target speaker's voice snippet and jointly separating all participating speakers by using a pool of additional speaker signals, which is known as speech separation using speaker inventory (SSUSI). However, all these systems ideally assume that the pre-enrolled speaker signals are available and are only evaluated on simple data configurations. In realistic multi-talker conversations, the speech signal contains a large proportion of non-overlapped regions, where we can derive robust speaker embedding of individual talkers. In this work, we adopt the SSUSI model in long recordings and propose a self-informed, clustering-based inventory forming scheme for long recording, where the speaker inventory is fully built from the input signal without the need for external speaker signals. Experiment results on simulated noisy reverberant long recording datasets show that the proposed method can significantly improve the separation performance across various conditions.

%% file: introduction.tex

Single-channel speech separation has been a challenging speech signal processing problem, and deep learning has provided advanced methods toward solving this problem \cite{hershey2016deep, yu2017permutation, luo2018speaker,han2019online,luo2019conv,wisdom2020unsupervised, jenrungrot2020cone}. In recent years, research that leverages additional speaker information has received increasing attention \cite{vzmolikova2019speakerbeam,delcroix2020improving,wang2018voicefilter,xiao2019single,wang2018deep,ochiai2019unified,wang2019speech}. We can categorize them into two main categories. The first category is informed speech extraction, which exploits an additional voice snippet of the target speaker to distinguish his/her speech from the mixture. SpeakerBeam \cite{vzmolikova2019speakerbeam,delcroix2020improving} derives a speaker embedding from an utterance of the target speaker by using a sequence summary network \cite{vesely2016sequence} and uses the embedding to guide an extraction network to extract the speaker of interest. VoiceFilter \cite{wang2018voicefilter} concatenates spectral features of the mixture with the d-vector \cite{variani2014deep} of a voice snippet to extract the target speaker. Xiao et al. \cite{xiao2019single} uses an attention mechanism to generate context-dependent biases for target speech extraction. Informed speech extraction naturally solves the permutation problem and unknown number of speakers. However, it has two limitations. Firstly, the computation cost is proportional to the number of speakers to be extracted, so in a multi-speaker conversation, the system needs to run multiple times to extract each speaker one by one. Most importantly, the extraction usually fails when the target speaker's biased information is not strong enough \cite{delcroix2020improving}.

The second category is speech separation using speaker inventory (SSUSI) \cite{wang2019speech}. The method employs a pool of additional enrollment utterances from a list of candidate speakers, from which profiles of relevant speakers involved in the mixture are first selected. Then the method fuses the selected profiles and the mixed speech to separate all speakers simultaneously. As multiple profiles are provided during separation, more substantial speaker discrimination can be expected, which yields better speech separation. The method can also employ permutation invariant training (PIT) \cite{yu2017permutation} to compensate for weak biased information and wrong selection.

Though with promising results reported in prior arts, both categories suffer from two issues. Firstly, as the separation performance heavily relies on the profile quality, when there is a severe acoustic mismatch between the mixed signal and the enrolled utterances, the effectiveness of speaker information could be largely degraded. Secondly, methods in both categories assume additional speaker information is available ahead of extraction or separation, which may be impractical in real scenarios. Wavesplit \cite{zeghidour2020wavesplit} uses clustering to infer source embeddings from the mixed signal and then uses them to guide speaker separation. However, the number of source embeddings must be fixed and identical to the speakers to be separated, limiting its application in a long recording with various speakers. Also, all the methods mentioned above mostly prove their successes on relative simple datasets, e.g., LibriMix \cite{cosentino2020librimix} that contains only anechoic speech, or WSJ0-2mix \cite{hershey2016deep} and its variants that contain pre-segmented speech utterances that are usually fully overlapped. These further blur the practicality of these methods as overlap in real conversation usually possess very different characteristics \cite{janin2003icsi,carletta2005ami,yoshioka2019advances,barker2018fifth}. 

In this paper, we address these problems on the continuous speech separation (CSS) task \cite{yoshioka2018multi,chen2020continuous}. CSS focuses on separating long recordings where the overall overlap ratio is low and the speaker activations are sparse. A large proportion of non-overlapped regions in the recording enables the derivation of robust features for the participants. We adopt the SSUSI in the CSS task and propose continuous SSUSI (CSSUSI), which constructs the speaker inventory from the mixed signal itself, instead of external speaker enrollments, by using speaker clustering methods. CSSUSI informs the separation network with relevant speaker profiles dynamically selected from the inventory to facilitate source separation at local regions. The outputs from local regions are then concatenated such that the output audio streams are continuous speech that do not contain any overlap. We create a more realistic dataset that simulates natural multi-talker conversations in conference rooms to test CSSUSI on the CSS task. Experimental results show that CSSUSI can successfully build a speaker inventory from the long speech mixture using the clustering-based method and take advantage of the global information to improve separation performance significantly.  

The rest of the paper is organized as follows. We introduce the SSUSI framework in Section~\ref{sec:s2}, describe the CSSUSI system for long recording in Section~\ref{sec:s3}, present the experiment configurations in Section~\ref{sec:exp}, analyze the experiment results in Section~\ref{sec:results}, and conclude the paper in Section~\ref{sec:con}.

%% file: ssusi.tex
We first overview the original SSUSI system \cite{wang2019speech}, which requires pre-enrolled speaker signals. A SSUSI system contains three modules: a speaker identification module, a speaker profile selection module, and a biased speech separation module. The speaker identification module is responsible for embedding extraction from both the speaker enrollments and input mixture. Embeddings of speaker enrollments are used for speaker inventory construction. The speaker profile selection module selects from the inventory the best-matched speaker profiles with the mixture embeddings. The selected profiles are then fed into the biased separation module to separate speakers in the mixture.

Since each speech segment is short (4s in this paper) and typically contains at most two speakers, we focus on two-speaker separation for each speech segment, and the model always generates two outputs. Moreover, we make several modifications to the original SSUSI architecture \cite{wang2019speech} for better performance.

\subsection{Speaker identification module}

The speaker identification module is used to construct the speaker inventory first. The inventory is a pool of $K$-dimensional speaker embeddings $\left\{\vec{e}^j\right\}_{j=1}^M, \vec{e}^j \in \mathbb{R}^{K}$, which are extracted from a collection of time-domain enrollment speech $\left\{\vec{a}^j\right\}_{j=1}^M, \vec{a}^j \in \mathbb{R}^{L_{a_j}}$, where $L_{a_j}$ is the temporal dimension of speech signal $\vec{a}^j$. $M$ is typically larger than the maximum number of speakers in the mixture to be separated. We also assume that each speaker only has one enrollment sentence. A speaker identification network, referred to as the \textit{SNet}, is applied for embedding extraction:
\begin{align}
    \label{eq:emb}
	\vec{E}^{j} = \rm{SNet}(\vec{a}^j)
\end{align}
where $\vec{E}^{j} \in \mathbb{R}^{T_{j}\times K}$ and $T_{j}$ is the temporal dimension of the embedding sequence. Here we simply use mean-pooling across the $T_{j}$ frames of $\vec{E}^{j}$ to obtain the single vector $\vec{e}^j \in \mathbb{R}^{K}$.

The mixture embeddings are directly extracted from the input mixture $\vec{y} \in \mathbb{R}^{T}$ with the temporal dimension T:
\begin{align}
    \label{eq:emb_mix}
	\vec{E}^{y} = \rm{SNet}(\vec{y})
\end{align}
where $\vec{E}^{y} \in \mathbb{R}^{T_{y}\times K}$ and $T_{y}$ is the temporal dimension of the mixture embeddings. 

\subsection{Speaker profile selection module}
\label{sec:profile}

The speaker profile selection module selects the relevant speaker profiles from the inventory that are best matched with the mixture embeddings $\vec{E}^{y}$ in equation \ref{eq:emb_mix}. The selection is performed by calculating the similarity between the mixture embeddings and items in the inventory, and two items with the highest similarity are selected. The similarity are calculated by applying the Softmax function on the dot-product between the mixture and inventory embeddings:
\begin{align}
\begin{split}
	\vec{d}_s^{y,j} &= \vec{e}_s^{y} \cdot \vec{e}^j \\
	\vec{w}_s^{y,j} &= \frac{\rm exp(\vec{d}_s^{y,j})}{\sum_{p=1}^M \rm exp(\vec{d}_s^{y,p})}
\end{split}
\end{align}
where $\vec{e}_s^{y}$ denotes $\vec{E}^{y}$ at temporal index $s$. We then calculate the average score $\vec{w}^{y,j}$ across the $T_{y}$ frames:
\begin{align}
	\vec{w}^{y,j} &= \frac{\sum_{s=1}^{T_{y}}\vec{w}_s^{y,j}}{T_{y}}
\end{align}
Two inventory items $\vec{e}^{p_1}$ and $\vec{e}^{p_2}$ are then selected according to the two highest scores in $\left\{\vec{w}^{y,j}\right\}_{j=1}^M$.

\begin{figure*}[!t]
    \centering
    \includegraphics[width=1.9\columnwidth]{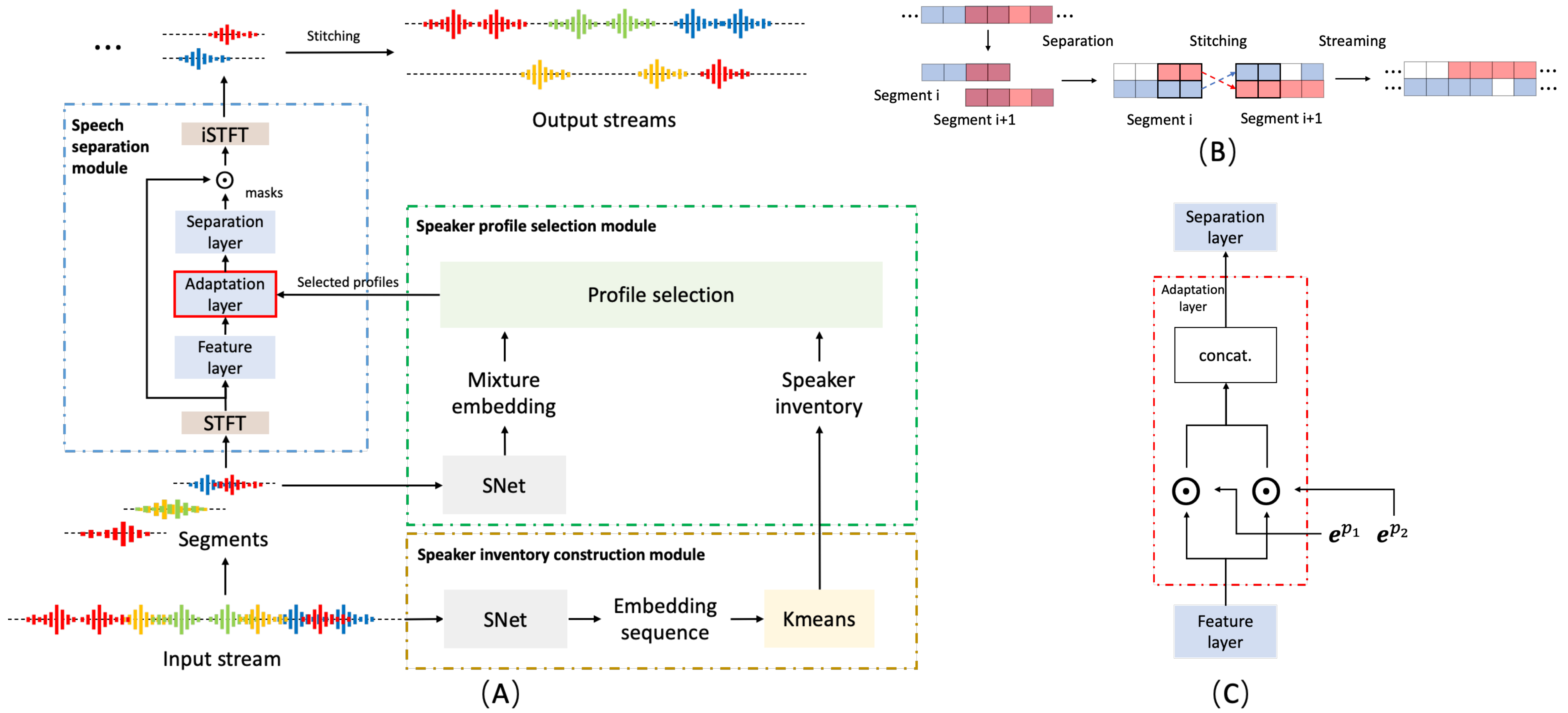}
    \caption{(A) The architecture of the proposed continuous speech separation using speaker inventory. The Speaker inventory construction module forms the speaker inventory from the long mixture by using Kmeans clustering; the long mixture is split into small segments, and the speaker profile selection module selects two relevant profiles from the inventory for each segment; the speech separation module fuses the selected speaker profiles into the system for source separation. (B) Multiplicative adaptation of the selected profiles $\vec{e}^{p_1}$ and $\vec{e}^{p_2}$. (C) Stitching procedure of adjacent segment outputs in a long recording.}
    \label{fig:arch}
\end{figure*}

\subsection{Biased speech separation module}
\label{sec:sep}

The biased speech separation module is adapted to the speech characteristics of the speakers selected from the inventory for biased source separation. The module contains three layers, a feature extraction layer, a profile adaptation layer, and a separation layer. Both feature extraction and separation layers are 2-layer BLSTM in this paper. Previous research \cite{vzmolikova2019speakerbeam} has shown that a multiplicative adaptation layer, i.e., multiplying the speaker embedding with the output of one of the middle layers of the network, is a simple yet effective way to realize adaptation, so we use the same method here. Given the two selected speaker profiles $\vec{e}^{p_1}$ and $\vec{e}^{p_2}$, two target-biased adaptation features are calculated by frame-level element-wise multiplication between the profiles and the output of the feature extraction layer:
\begin{align}
    \vec{a}_l^{p_1} &= \vec{b}_l \odot \vec{e}^{p_1} \\
    \vec{a}_l^{p_2} &= \vec{b}_l \odot \vec{e}^{p_2} 
\end{align}
where $\vec{b}_l \in \mathbb{R}^{K}$ denotes the output of the feature layer, $l$ denotes the frame index, and $\odot$ denotes the element-wise multiplication. The two target-biased features are then concatenated:
\begin{align}
    \vec{A} = \rm concat([\vec{A}^{p_1}, \vec{A}^{p_2}])
    \label{eqn:cat}
\end{align}
where $\vec{A}^{p_1} = [\vec{a}_1^{p_1}, \ldots, \vec{a}_L^{p_1}] \in \mathbb{R}^{L\times K}$, $\vec{A}^{p_2} = [\vec{a}_1^{p_2}, \ldots, \vec{a}_L^{p_2}] \in \mathbb{R}^{L\times K}$, and $\vec{A} \in \mathbb{R}^{L\times 2K}$. The separation layer takes $\vec{A}$ as the input and estimates two time-frequency (T-F) masks $\vec{M}^1, \vec{M}^2 \in \mathbb{R}^{L\times F}$.

%% file: cssusi.tex
SSUSI assumes that pre-recorded utterances of all speakers are available for the speaker inventory construction. However, such an assumption may not be realistic, especially for unseen speakers or meeting scenarios where the collection of pre-recorded speech from the participants is not feasible. 

Continuous speech separation (CSS) aims at estimating the individual target signals from a continuous mixed signal which is usually a hours long signal and contains both overlapped and non-overlap speech, but the overlap ratio is low. So, single-speaker regions can be exploited to derive robust acoustic characteristics of participating speakers without the need for external utterances, which makes the self-informed speaker inventory construction possible. This section introduces how we adopt SSUSI in the CSS task and eliminate the need for pre-recorded speech by using a clustering method.

Figure~\ref{fig:arch} (A) shows the overall flowchart of the continuous SSUSI (CSSUSI) framework. The main difference between CSSUSI and the original SSUSI is the construction of the speaker inventory.  Original SSUSI applies the speaker identification module on extra enrollment utterances, whereas CSSUSI first splits the mixture recording $\vec{y}$ into $B$ small chunks, and directly extracts the mixture embeddings $\left\{\vec{e}_b^y\right\}_{b=1}^B$, where $\vec{e}_b^y \in \mathbb{R}^{K}$ denotes the embedding vector in chunk b. Then, CSSUSI applies Kmeans clustering on $\left\{\vec{e}_b^y\right\}_{b=1}^B$ to form $M$ clusters, and the cluster centroids form the speaker inventory. In Section~\ref{sec:results} we will show that the separation performance is insensitive to the choice of $M$ as long as $M$ is no smaller than the actual number of active speakers in the recording.

CSUSSI uniformly segments the mixture recording and exploits the inventory to facilitate source separation in each segment. Except for the self-informed speaker inventory, CSSUSI uses the same speaker profile selection and biased speech separation methods as introduced in Section \ref{sec:profile} and Section \ref{sec:sep}, respectively. To stitch the outputs from different segments to form output streams where each stream only contains non-overlapped speakers, the similarity between the overlapped regions in adjacent blocks determines the pair of segments to be stitched. Figure~\ref{fig:arch} (C) shows the stitching procedure of adjacent segment outputs.

%% file: experiments.tex
\begin{table*}[!ht]
	\scriptsize
	\centering
	\caption{SNR (dB) on eight-speaker long recordings (segment-wise evaluation). The performance on different overlap ratios is reported.}
	\begin{tabular}{c|c|cccccc}
		\thline
        \multirow{2}{*}{\thead{Method}} & \multirow{2}{*}{\thead{Speaker enrollment}} & \multicolumn{6}{c}{\thead{Overlap ratio in $\%$}} \\
        & & 0 & 0-25 & 25-50 & 50-75 & 75-100 & Average \\
        \thline
        \multirow{1}{*}{Unprocessed}
        & - & 8.6 & -9.7 & -1.2 & -0.9 & -0.7 & -0.1 \\
        \hline
        \multirow{1}{*}{BLSTM}
        & - & 15.5 & 8.0 & 8.6 & 7.5 & 6.9 & 10.6 \\
        \hline
        \multirow{3}{*}{SSUSI}
        & Two wrong profiles & 15.2 & 7.1 & 8.4 & 7.8 & 7.1 & 10.3 \\
        & One correct and one wrong profiles & 15.4 & 7.8 & 9.0 & 8.2 & 7.6 & 10.7 \\
        & Two correct profiles & 15.9 & 9.5 & 10.6 & 9.4 & 8.7 & 11.9 \\
        & Selected profiles & 15.7 & 8.8 & 10.0 & 9.0 & 8.3 & 11.5 \\
        \thline
    \end{tabular}
    \label{tab:t1}
\end{table*}

\subsection{Dataset}
\label{sec:data}

In our training set, we randomly generate 3000 rooms. The length and width of the rooms are randomly sampled between 5 and 12 meters, and the height is randomly sampled between 2.5 and 4.5 meters. A microphone is randomly placed in the room, and its location is constrained to be within 2 meters of the room center. The height of the microphone is randomly sampled between 0.4 and 1.2 meters. We randomly sample 10 speakers from the LibriSpeech corpus \cite{panayotov2015librispeech} for each room. All the speakers are at least 0.5 meters away from the room walls and the height of the speakers are between 1 and 2 meters. The reverberation time is uniformly sampled between 0.1 and 0.5 seconds. We randomly choose 2 speakers as relevant speakers and arrange them according to one of the four following patterns:
\begin{enumerate}
    \item \textit{Inclusive}: one speaker talks a short period while the other one is talking.
    \item \textit{Sequential}: one talks after the other one finishes talking.
    \item \textit{Fully-overlapped}: two speakers always talk simultaneously.
    \item \textit{Partially-overlapped}: two speakers talk together only in a certain period.
\end{enumerate}
The frequencies for the four patterns are 10\%, 20\%, 35\%, and 35\%, respectively. The minimal length of the overlapped periods in inclusive and partially-overlapped patterns is set to 1 second. The maximal length of the silent periods between the two speakers in the sequential pattern is 0.5 second. Moreover, to generate single-speaker utterances, there is a 0.1 probability that one of the speakers is muted in each pattern. We use the remaining 8 speakers as the irrelevant speakers that will not appear in the mixture. Each of the room configurations is used for 8 times. The mixture length is 4 seconds. So, the total training time is $3000\times 8\times 4s = 26.7\ \text{hours}$.  For both the relevant and irrelevant speakers, a 10-second utterance is sampled to form the speaker inventory. All speech signals are single-channel and sampled at 16 kHz. Gaussian noise with SNR randomly chosen between 0 and 20 dB is added into the mixture. 

In our testing set, we set three configurations: $60$-second mixture containing 2 speakers, $150$-second mixture containing 5 speakers, and $240$-second mixture containing 8 speakers. We generate 300 recordings for each configuration. The overall overlap ratio of each recording is 30$\%$ complying with natural conversion \cite{ccetin2006analysis}.

\subsection{Implementation details}

All models contain 4 bidirectional LSTM (BLSTM) layers with 600 hidden units in each direction. In the CSSUSI models, the speaker identification module adopts the similar design in \cite{zhou2019cnn}, and the module is pretrained on the VoxCeleb2 dataset \cite{Chung2018vox2} and achieves 2.04$\%$ equal error rate on the VoxCeleb1 test set \cite{Nagrani2017vox1}. The module extracts 128-dimensional speaker embeddings for every 1.2-second (30-frame) segment. 
We use SNR as training objective \cite{han2020real} and Adam \cite{kingma2014adam} as the optimizer with initial learning rate of 0.001. The learning rate is decayed by 0.98 for every two epochs.



%% file: results.tex
\begin{table}[!ht]
	\scriptsize
	\centering
	\caption{SNR (dB) on long recordings with different configurations (segment-wise evaluation).} 
	\begin{tabular}{c|c|c|c|c}
		\thline
        Speaker number & Method & External utterances & Clusters & Avg. \\
        \thline
        \multirow{6}{*}{2 speakers} 
        & Unprocessed & - & - & 1.6 \\
        \cline{2-5}
        & BLSTM & - & - & 11.2 \\
        \cline{2-5}
        & SSUSI & 2 & No & 12.2 \\
        \cline{2-5}
        & \multirow{3}{*}{\thead{CSSUSI}} 
        & \multirow{3}{*}{\thead{No}} 
        &     2  & 12.1 \\
        & & & 3 & 11.9 \\
        & & & 4 & 11.9 \\
        \thline
        \multirow{7}{*}{5 speakers} &
        Unprocessed & - & - & 0 \\
        \cline{2-5}
        & BLSTM & - & - & 10.6 \\
        \cline{2-5}
        & SSUSI & 5 & No & 11.5 \\
        \cline{2-5}
        & \multirow{4}{*}{\thead{CSSUSI}}
        & \multirow{4}{*}{\thead{No}} 
        &     3  & 10.9 \\
        & & & 5 & 11.3 \\
        & & & 8 & 11.2 \\
        & & & 10 & 11.2 \\
        \thline
        \multirow{7}{*}{8 speakers} &
        Unprocessed & - & - & -0.1 \\
        \cline{2-5}
        & BLSTM & - & - & 10.6 \\
        \cline{2-5}
        & SSUSI & 8 & No & 11.5 \\
        \cline{2-5}
        & \multirow{4}{*}{\thead{CSSUSI}} 
        & \multirow{4}{*}{\thead{No}} 
        &        5 & 11.0 \\
        & & &  8 & 11.3 \\
        & & & 12 & 11.3 \\
        & & & 16 & 11.2 \\
        \thline
    \end{tabular}
    \label{tab:t2}
\end{table}

Table~\ref{tab:t1} compares different models on 4-second segments of eight-speaker recordings. The inventory contains eight speakers' profiles that are derived from eight external utterances. SSUSI achieves leading performance on all levels of overlap ratios when two correct speaker profiles are used; however, the performance of SSUSI drops greatly with two wrong speaker profiles randomly chosen from the 8 irrelevant speakers, which indicates that performance gain obtained by SSUSI mainly comes from leveraging the target speaker information. We also notice that the performance of SSUSI with two wrong profiles is only slightly worse than the baseline BLSTM, and when only one correct speaker profile is enrolled, SSUSI can still outperform the baseline model, which proves that PIT can compensate for wrong selection and the separation module is robust to adaptation features. When the speaker profiles are selected by the profile selection module, the SSUSI model performs slightly better on the non-overlapped mixtures (overlap ratio is 0) but much better on the overlapped mixtures at all overlap ratios. This confirms the effectiveness of the SSUSI framework on improving separation performance across various settings, which is consistent with the observations in \cite{wang2019speech} that conducted experiments on Librispeech although the model architectures are different.

Table~\ref{tab:t2} compares CSSUSI with different clusters on recordings with different number of speakers. Since the number of participating speakers in a meeting may be unknown, we intend to do over-clustering, i.e., setting the number of clusters greater than the number of speakers in a meeting. Table~\ref{tab:t2} compares CSSUSI with different clustering settings. The performance of CSSUSI is almost identical once the number of clusters is not fewer than the number of speakers. Over-clustering has very little impact on the performance as it ensures each speaker possesses at least one cluster center. Some extra clusters may represent acoustic characteristics of overlapped regions, which will be regarded as irrelevant profiles during profile selection. We can see that CSSUSI outperforms the baseline model BLSTM on all configurations. As we conclude from Table~\ref{tab:t1}, the performance gain is achieved via leveraging relevant speakers' information. So the performance gain from CSUSSI suggests the successful construction of the speaker inventory from the mixture itself and effective utilization of speaker information. Furthermore, we compare CSSUSI with SSUSI that derives speaker profiles from external utterances that contain only a single speaker in each utterance. CSSUSI sacrifices very little performance but does not require external utterances, which shows CSSUSI is a better model than SSUSI for long recording speech separation.

Table~\ref{tab:t3} compares utterance-wise separation performance. After segments are stitched, each complete utterance is extracted from the output streams by using ground-truth segmentation information, i.e., onset and offset of each utterance. We find that CSSUSI surpasses the baseline in all configurations by a large margin, which further proves the strength of CSSUSI in the long recordings.

\begin{table}[!t]
	\scriptsize
	\centering
	\caption{Utterance-level evaluation. SI-SDR(dB) is reported.}
	\label{tab:ssusi}
	\begin{tabular}{c|c|ccc}
		\thline
		Method & Need external utterances? & 2 spk & 5 spk & 8spk \\
		\thline
		Unprocessed & - & 6.0 & 4.5 & 4.3 \\
		\hline
		BLSTM  & No & 11.7 & 10.8 & 10.6 \\
		\hline
		SSUSI  & Yes & 13.2 & 12.0 & 11.7 \\
		\hline
		CSSUSI & No & 13.1 & 11.9 & 11.7 \\
        \thline
    \end{tabular}
    \label{tab:t3}

\end{table}

%% file: conclusion.tex
In this paper, we investigated continuous speech separation using speaker inventory for long multi-talker recordings.  In the CSS task, we made use of the fact that long recording, in general, contains a large proportion of non-overlapped regions and proposed continuous SSUSI (CSSUSI) that extracted speaker embeddings from the long recordings and performed ``over-clustering'' on the embeddings to construct the self-informed speaker inventory. CSSUSI overcomes the limitation of the original SSUSI that required external enrollments. Experiments on a simulated noisy reverberant dataset showed that CSSUSI significantly outperformed the baseline models across various conditions. Future works include extending the CSSUSI system into real-world recordings, designing a block-online system instead of an offline system, and investigate better model architectures.